\documentclass[aps,prd,onecolumn,groupedaddress,showpacs,nofootinbib,amssymb]{revtex4}
\usepackage[latin1]{inputenc}
\usepackage{graphicx}
\usepackage[english]{babel}

\usepackage{amsmath}
\usepackage{amssymb}
\usepackage{amsfonts}
\usepackage{colordvi}
\usepackage{psfrag}
\usepackage{color}

\begin{document}
\def\pp{{\, \mid \hskip -1.5mm =}}
\def\cL{{\cal L}}
\def\be{\begin{equation}}
\def\ee{\end{equation}}
\def\bea{\begin{eqnarray}}
\def\eea{\end{eqnarray}}
\def\beq{\begin{eqnarray}}
\def\eeq{\end{eqnarray}}
\def\tr{{\rm tr}\, }
\def\nn{\nonumber \\}
\def\e{{\rm e}}
\def\de{\partial}
\def\lm{{\ell m}}

\title{Axially symmetric solutions in $f(R)$-gravity}

\author{Salvatore Capozziello$^1$, Mariafelicia De Laurentis$^1$, Arturo Stabile$^2$}
\affiliation{\it $^1$Dipartimento di Scienze Fisiche, Università
di Napoli {}``Federico II'', INFN Sez. di Napoli, Compl. Univ. di
Monte S. Angelo, Edificio G, Via Cinthia, I-80126, Napoli, Italy\\ $^2$ Dipartimento di Ingegneria,
Università del Sannio,
Benevento, C.so Garibaldi
107, I-80125 Benevento, Italy}
\date{\today}

\begin{abstract}
Axially symmetric solutions for $f(R)$-gravity can be  derived
starting from  exact spherically symmetric solutions achieved by Noether symmetries. The method takes advantage
of a complex coordinate transformation  previously developed by Newman and Janis in General Relativity.
 An example is worked out to show the general validity of the approach. The physical properties of the solution are also considered.
\end{abstract}


\maketitle
\section{Introduction}
\label{uno}

The issue to extend General Relativity (GR) to alternative theories of gravity has recently become dramatically urgent due to  the missing matter problem at all astrophysical scales and the accelerating behavior of cosmic fluid, detected by SuperNovae Ia used as standard candles. Up to now, no final answer on new particles has been given at fundamental level so Dark Energy and Dark Matter  constitute a puzzle to be solved in order to achieve a self-consistent picture of the observed Universe. $f(R)$-gravity, where $f(R)$ is a generic function of the Ricci scalar $R$,  comes into the game as a straightforward extension of GR where further geometrical degrees of freedom are considered instead of searching for new material ingredients \cite{f(R)-cosmo}. From an epistemological  point of view, the action of gravity  is not selected {\it a priori}, but it could be "reconstructed", in principle, by matching consistently the observations \cite{Capozz2, GRGrev,FoP}. This approach can be adopted considering any function of the curvature invariants as $R_{\mu\nu}R^{\mu\nu}$, $R\Box R$ and so on.

From a genuine mathematical point of view, alternative theories of gravity pose the problem to recover or extend the well-established results of GR as the initial value problem \cite{cauchy},  the stability of solutions  and, in particular, the issue of finding out new solutions.  As it is well known, beside cosmological solutions, spherically and axially symmetric solutions play a fundamental role in several astrophysical problems ranging from black holes to active galactic nuclei. Alternative gravities, to be consistent with  results of GR, should comprise solutions like Schwarzschild and Kerr ones but present, in general,  new solutions that could be physically interesting. Due to this reason, methods to find out exact and approximate solutions are particularly relevant in order to check if observations can be framed in Extended Theories of Gravity \cite{noether}.

Recently, the interest in spherically symmetric solutions of $f(R)$-gravity is growing up. In  \cite{Multamaki},  solutions in vacuum
have been found considering relations among functions that define the spherical metric or
imposing a costant Ricci curvatue scalar. The authors have reconstructed  the form of some $f(R)$-models,
discussing their physical relevance. In \cite{Multamaki1}, the
same authors have discussed static spherically symmetric  solutions,  in presence of perfect
fluid matter, adopting the metric formalism. They have
shown that a given matter distribution is not capable of globally determining
the functional form of $f(R)$. Others authors have discussed  in
details the spherical symmetry  of $f(R)$-gravity considering also
the relations with the weak field limit.  Exact solutions are
obtained for constant Ricci curvature scalar and for Ricci scalar
depending on the radial coordinate. In particular, it can be considered
how to obtain results  consistent with GR assuming the well-known post-Newtonian and post-Minkowskian limits as consistency checks.
\cite{arturo}.

In this paper, we want to seek for a general metod to find out
axially symmetric solutions  by performing a complex coordinate transformation
on the spherical metrics. Since the discovery of the Kerr solution \cite{pap:kerr},  many attempts have been made to find a
physically reasonable interior matter distribution that may be
considered as its source. For a review on these approaches
see \cite{pap:dmm, pap:ak}. Though much progress has been made, results have
been generally disappointing. As far as we know,  nobody  has
obtained a physically satisfactory interior solution. This seems
surprising given the success of matching internal spherically symmetric solutions to the Schwarzschild metric. The problem is
 that the loss of a degree of symmetry makes the
derivation of analytic results  much more difficult.  Severe
restrictions are placed on  the interior metric by maintaining that it
must be joined smoothly to the external axially symmetric metric.
Further restrictions are placed on the  interior solutions to ensure that they correspond to
physical objects.

Furthermore
since the axially symmetric metric has no radiation field associated with it,
its source should  be also non-radiating. This places even
further constraints on the structure of the interior
solution  \cite{bk:dk}. Given the strenuous nature of these limiting
conditions, it  is not surprising to learn that  no
satisfactory solution to the problem of finding  sources
for the Kerr metric has been obtained. In general,  the failure is  due to  internal
structures whose physical properties are unknown. This shortcoming makes hard to find consistent
 boundary conditions.

Newman and Janis
showed that it  is possible to  obtain an  axially symmetric  solution (like the
Kerr metric) by making an elementary complex transformation on the
Schwarzschild solution~\cite{pap:nj1}. This same method has been
used to obtain a new stationary and axially symmetric
solution known as the
Kerr-Newman metric~\cite{pap:nj2}. The Kerr-Newman space-time is
associated to the exterior geometry of a rotating massive and
charged black-hole. For a  review on the Newman-Janis method
 to obtain both the Kerr and Kerr-Newman metrics
see~\cite{bk:ier}.

By means of very elegant
mathematical arguments, Schiffer et al.~\cite{pap:mms} have  given a rigorous
proof to show how the Kerr metric can be derived starting from a complex
transformation on the Schwarzschild solution. We will not go
into the details of this demonstration, but point out  that
 the proof  relies on two main assumptions. The first  is that the metric
belongs to the same algebraic class of the Kerr-Newman solution,
namely the Kerr-Schild class \cite{pap:gcd}. The second assumption is
that the metric corresponds to an empty solution of the Einstein
field equations. In the case we are going to study, these assumptions are not
considered and hence the proof  in \cite{pap:mms} is not applicable. It is clear, by the
generation of the Kerr-Newman metric, that all the components of
the stress-energy tensor need to be non-zero for the Newman-Janis method to be
successful. In fact, G\"{u}rses and G\"{u}rsey, in
1975~\cite{pap:mg}, showed that if a metric can be written in the Kerr-Schild
form, then a complex transformation ``is allowed in General
Relativity.'' In this paper,  we will show that  such a transformation
can be  extended to $f(R)$-gravity.

The outline of this paper is as follows. In the Sec.\ref{due}, we
introduce  the $f(R)$-gravity action,  the field
equations and  give some general remarks
on spherical symmetry. In Sec ~\ref{noe},  we give a summary of  the Noether Simmetry Approach  \cite{noether}
and find  some spherically symmetric exact solutions for $f(R)$-gravity.
In Sec ~\ref{tre},
 we review the Newman-Janis method  to obtain axially symmetric solutions
starting from spherically symmetric ones. The resulting metric is written in terms of
two arbitrary functions.  A further suitable   coordinate
transformation allows to write the  metric  in the so called
Boyer-Lindquist  coordinates. Such a transformation makes the physical
interpretation much clearer and reduces the amount of algebra
required to calculate the  metric properties.
In Sec.\ref{quattro}, the Newman-Janis method  is applied to the spherically symmetric exact solution, previously derived
 by the Noether Symmetry, and an axially symmetric exact
solution is obtained. This result shows  that the Newman-Janis method works also in $f(R)$-gravity.
A physical application of the result is discussed in Sec \ref{cinque}. Discussion ans concluding remarks are drawn in Sec. \ref{sei}.

\section{Spherical symmetry in  $f(R)$-gravity }
\label{due}

Let us consider an analytic function $f(R)$ of the Ricci scalar
$R$ in four dimensions. The variational principle for this action is:

\begin{equation}\label{fRaction}
\delta\int
d^4x\sqrt{-g}\biggl[f(R)+\mathcal{X}\mathcal{L}_m\biggr]\,=\,0
\end{equation}
where ${\displaystyle \mathcal{X}=\frac{8\pi G}{c^4}}$,
$\mathcal{L}_m$ is the standard matter Lagrangian and $g$ is the
determinant of the metric\footnote{We are adopting  the convention
$R_{\mu\nu}={R^\rho}_{\mu\rho\nu}$ for the Ricci tensor and
${R^\alpha}_{\beta\mu\nu}=\Gamma^\alpha_{\beta\nu,\mu}-...$, for
the Riemann tensor. Connections are Levi-Civita \,:
\begin{equation}
\Gamma^\mu_{\alpha\beta}=\frac{1}{2}g^{\mu\rho}(g_{\alpha\rho,\beta}+g_{\beta\rho,\alpha}-g_{\alpha\beta,\rho})\,.\nonumber\\
\end{equation}}.

By varying  with respect to the metric, we obtain the field
equations \footnote{It is possible to take into account also the
Palatini approach in which  the metric $g$ and the connection
$\Gamma$  are considered independent variables (see for example
\cite{palatini}). Here we will consider the Levi-Civita connection
and will use the metric approach. See \cite{GRGrev,ACCF} for a
detailed comparison between the two pictures.}

\begin{equation}\label{HOEQ}
\left\{\begin{array}{ll}H_{\mu\nu}=f'(R)R_{\mu\nu}-\frac{1}{2}f(R)g_{\mu\nu}-f'(R)_{;\mu\nu}+g_{\mu\nu}\Box
f'(R)\,=\,\mathcal{X}T_{\mu\nu}\\\\H\,=\,g^{\rho\sigma}H_{\rho\sigma}=3\Box
f'(R)+f'(R)R-2f(R)\,=\,\mathcal{X}T\end{array}\right.
\end{equation}
where
$T_{\mu\nu}\,=\,\displaystyle\frac{-2}{\sqrt{-g}}\frac{\delta(\sqrt{-g}\mathcal{L}_m)}{\delta
g^{\mu\nu}}$ is the energy-momentum tensor of standard fluid matter and the second equation is the trace. The
most general spherically symmetric solution can be written as
follows\,:

\begin{equation}\label{me0}
ds^2\,=\,m_1(t',r')dt'^2+m_2(t',r')dr'^2+m_3(t',r')dt'dr'+m_4(t',r')d\Omega\,,
\end{equation}
where $m_i$ are functions of the radius $r'$ and of the time $t'$.
 $d\Omega$ is the solid angle. We can consider a
coordinate transformation that maps the metric (\ref{me0}) in a
new one where the off\,-\,diagonal term vanishes and
$m_4(t',r')\,=\,-r^2$, that is\footnote{This condition allows to
obtain the standard definition of the circumference with  radius
$r$.}\,:

\begin{equation}\label{me}
ds^2\,=\,g_{tt}(t,r)dt^2-g_{rr}(t,r)dr^2-r^2d\Omega\,.
\end{equation}
This  expression can be considered, without loss of generality, as
the most general definition of a spherically symmetric metric
compatible with a pseudo\,-\,Riemannian manifold  without torsion.
Actually, by inserting this metric into the field Eqs.
(\ref{HOEQ}), one obtains\,:

\begin{equation}\label{fe4}
\left\{\begin{array}{ll}f'(R)R_{\mu\nu}-\frac{1}{2}f(R)g_{\mu\nu}+\mathcal{H}_{\mu\nu}\,=\,\mathcal{X}T_{\mu\nu}\\\\
f'(R)R-2f(R)+\mathcal{H}\,=\,\mathcal{X}T\end{array}\right.
\end{equation}
where the two quantities $\mathcal{H}_{\mu\nu}$ and $\mathcal{H}$
read\,:
\begin{eqnarray}\label{highterms1}
\mathcal{H}_{\mu\nu}\,=\,-f''(R)\biggl\{R_{,\mu\nu}-\Gamma^t_{\mu\nu}R_{,t}-\Gamma^r_{\mu\nu}R_{,r}-
g_{\mu\nu}\biggl[\biggl({g^{tt}}_{,t}+g^{tt}
\left(\ln\sqrt{-g}\right)_{,t}\biggr)R_{,t}+\biggl({g^{rr}}_{,r}+g^{rr}\left(\ln\sqrt{-g}\right)_{,r}\biggr)R_{,r}+\nonumber\\\nonumber\\+g^{tt}R_{,tt}
+g^{rr}R_{,rr}\biggr]\biggr\}-f'''(R)\biggl[R_{,\mu}R_{,\nu}-g_{\mu\nu}\biggl(g^{tt}{R_{,t}}^2+g^{rr}
{R_{,r}}^2\biggr)\biggr]
\\\nonumber\\\label{highterms2}
\mathcal{H}\,=\,g^{\sigma\tau}\mathcal{H}_{\sigma\tau}\,=\,3f''(R)\biggl[\biggl({g^{tt}}_{,t}+g^{tt}
\left(\ln\sqrt{-g}\right)_{,t}\biggr)R_{,t}+\biggl({g^{rr}}_{,r}+g^{rr}\left(\ln\sqrt{-g}\right)_{,r}\biggr)R_{,r}+g^{tt}R_{,tt}
+g^{rr}R_{,rr}\biggr]+\nonumber\\\nonumber\\+
3f'''(R)\biggl[g^{tt}{R_{,t}}^2+g^{rr}{R_{,r}}^2\biggr]\,.
\end{eqnarray}
Our task is now to find out
exact spherically symmetric solutions.

In the case of time-independent metric, i.e., $g_{tt}\,=\,a(r)$
and $g_{rr}\,=\,b(r)$, the Ricci scalar  can
be recast as a Bernoulli equation of index two with respect to
the metric potential $b(r)$\, (see \cite{arturo} for details):

\begin{eqnarray}\label{eqric}
b'(r)+\biggl\{\frac{r^2a'(r)^2-4a(r)^2-2ra(r)[2a(r)'+ra(r)'']}{ra(r)[4a(r)
+ra'(r)]}\biggr\}b(r)+\biggl\{\frac{2a(r)}{r}\biggl[\frac{2+r^2R(r)}{4a(r)+ra'(r)}\biggr]\biggr\}b(r)^2\,=\,0\,.
\end{eqnarray}
where $R\,=\,R(r)$ is the Ricci scalar. A general solution of
(\ref{eqric}) is:

\begin{equation}\label{gensol}
b(r)\,=\,\frac{\exp[-\int dr\,h(r)]}{K+\int dr\,l(r)\,\exp[-\int
dr\,h(r)]}\,,
\end{equation}
where $K$ is an integration constant while $h(r)$ and $l(r)$ are
 two functions that, according to Eq.(\ref{eqric}), define the coefficients of
the quadratic and the linear term with respect to $b(r)$
\cite{bernoulli}. We can fix  $l(r)\,=\,0$; this choice allows to find out
solutions with a Ricci scalar scaling as ${\displaystyle
-\frac{2}{r^2}}$ in term of the radial coordinate. On the other
hand, it is not possible to have $h(r)\,=\,0$ since, otherwise, we
 get imaginary solutions. A particular consideration deserves
the limit $r\rightarrow\infty$. In order to achieve a
gravitational potential $b(r)$ with the correct Minkowski limit,
both $h(r)$ and $l(r)$ have to go to zero at infinity, provided that the
quantity $r^2R(r)$ turns out to be constant: this result implies
$b'(r)=0$, and, finally, also the metric potential $b(r)$ has  a
correct Minkowski limit.

In general,  if we ask for the asymptotic flatness of the metric
as a feature of the theory,  the Ricci scalar has to evolve to
infinity as $r^{-n}$ with $n\geqslant 2$. Formally, it has to be:

\begin{equation}\label{condricc}
\lim_{r\rightarrow\infty}r^2R(r)\,=\,r^{-n}\,,
\end{equation}
with $n\in\mathbb{N}$. Any other behavior of the Ricci scalar
could affect the requirement to achieve a correct asymptotic flatness.

The case of constant curvature is equivalent to GR with a
cosmological constant and the solution is time independent. This
result is well known (see, for example, \cite{barrottew}) but we
report, for the sake of completeness, some considerations related
with it in order to deal with more general cases where a radial
dependence for the Ricci scalar is supposed.  If the scalar curvature is constant
($R\,=\,R_0$),  field Eqs.(\ref{fe4}), being
$\mathcal{H}_{\mu\nu}\,=\,0$,  reduce to:

\begin{equation}\label{fe2}
\left\{\begin{array}{ll}f'_0R_{\mu\nu}-\frac{1}{2}f_0g_{\mu\nu}\,=\,\mathcal{X}T_{\mu\nu}\\\\
f'_0R_0-2f_0\,=\,\mathcal{X}T\end{array}\right.
\end{equation}
where $f(R_0)=f_0$, $f'(R_0)=f'_0$. A general solution, when one considers a stress-energy tensor
of perfect-fluid $T_{\mu\nu}\,=\,(\rho+p)u_\mu u_\nu-pg_{\mu\nu}$,
is

\begin{equation}
ds^2\,=\,\biggl(1+\frac{k_1}{r}+\frac{q\mathcal{X}\rho-\lambda}{3}r^2\biggr)dt^2-\frac{dr^2}{1+\frac{k_1}{r}+
\frac{q\mathcal{X}\rho-\lambda}{3}r^2}-r^2d\Omega\,.
\end{equation}
when $p\,=\,-\rho$, $\lambda=-\frac{f_0}{2f'_0}$ and
$q^{-1}=f'_0$. This result means that any $f(R)$-model, in the
case of constant curvature, exhibits solutions
with de Sitter-like behavior. This is
one of the reasons why the dark energy issue can be addressed using
these theories \cite{f(R)-cosmo}.

If $f(R)$ is analytic, it is possible to write the series:
\begin{equation}\label{f}
f(R)\,=\,\Lambda+\Psi_0R+\Psi(R)\,,
\end{equation}
where $\Psi_0$ is a coupling constant, $\Lambda$ plays the role of
the cosmological constant and $\Psi(R)$ is a generic analytic
function of $R$ satisfying the condition

\begin{equation}\label{psi}
\lim_{R\rightarrow 0}R^{-2}\Psi(R)\,=\,\Psi_1\,,
\end{equation}
where $\Psi_1$ is a constant. If we neglect the cosmological
constant $\Lambda$ and $\Psi_0$ is set to zero, we obtain a new
class of theories which, in the limit $R\rightarrow{0}$, does not
reproduce GR (from  Eq.(\ref{psi}), we have $\lim_{R\rightarrow 0}
f(R)\sim R^2$). In such a case, analyzing the whole set of
Eqs.(\ref{fe2}), one can observe that both zero and constant $\neq
0$ curvature solutions are possible. In particular, if
$R\,=\,R_0\,=\,0$ field equations are solved for any form of
gravitational potential entering the spherically symmetric
background,  provided that the Bernoulli Eq. (\ref{eqric}),
relating these functions, is fulfilled for the particular case
$R(r)=0$. The  solutions are thus defined by the relation

\begin{equation}\label{gensol0}
b(r)\,=\,\frac{\exp[-\int
dr\,h(r)]}{K+4\int\frac{dr\,a(r)\,\exp[-\int
dr\,h(r)]}{r[a(r)+ra'(r)]}}\,,
\end{equation}
being $g_{tt}(t,r)=b(r)$ from Eq.(\ref{me}). In \cite{arturo},
some examples of $f(R)$-models admitting
solutions with constant$\neq 0$ or null scalar curvature are discussed.

\section{The Noether Symmetry Approach and the spherical symmetry }
\label{noe}

Besides spherically symmetric solutions with constant curvature scalar, also solutions with the Ricci scalar depending on radial coordinate $r$ are possible in $f(R)$-gravity \cite{arturo}. Furthermore, spherically symmetric solutions
 can be achieved  starting from a
point-like $f(R)$-Lagrangian  \cite{noether}. Such a Lagrangian can be obtained by
imposing the spherical symmetry  directly in the   action
(\ref{fRaction}). As a consequence, the infinite number of degrees
of freedom of the original field theory will be reduced to a
finite number. The technique is based on the choice of a suitable
Lagrange multiplier defined by assuming the Ricci scalar, argument
of the function $f(R)$ in spherical symmetry.

Starting from the above considerations, a static
spherically symmetric metric can be expressed as

\begin{equation}\label{me2}
{ds}^2=A(r){dt}^2-B(r){dr}^2-M(r)d\Omega\,,
\end{equation}
and then the point-like $f(R)$ Lagrangian\footnote{Obviously, the above choices are recovered for $A(r)=a(r)$, $B(r)=b(r)$, and $M(r)=r^2$. Here we deal with $A,B,M$ as a set of coordinates in a configuration space.}  is

\begin{eqnarray}\label{lag2}
\mathcal{L}=-\frac{A^{1/2}f'}{2MB^{1/2}}{M'}^2-\frac{f'}{A^{1/2}B^{1/2}}A'M'-\frac{Mf''}{A^{1/2}B^{1/2}}A'R'-\frac{2A^{1/2}f''}{B^{1/2}}R'M'-A^{1/2}B^{1/2}[(2+MR)f'-Mf]\,,
\end{eqnarray}
which is canonical since only the configuration variables and
their first order derivatives with respect to the radial coordinate $r$ are present. Details of calculations are in \cite{noether}.
 Eq.
(\ref{lag2}) can be recast in a more compact form introducing the
matrix representation\,:
\begin{equation}\label{la}
\mathcal{L}={\underline{q}'}^t\hat{T}\underline{q}'+V
\end{equation}
where $\underline{q}=(A,B,M,R)$ and $\underline{q}'=(A',B',M',R')$
are the generalized positions and velocities associated to
$\mathcal{L}$.  It is easy to check
the complete analogy between the field equation approach and
point-like Lagrangian approach \cite{noether}.

In order to find out solutions for the  Lagrangian (\ref{lag2}),
we can search for symmetries related to cyclic variables and then
reduce dynamics. This approach allows, in principle, to select
$f(R)$-gravity models compatible with spherical symmetry. As a
general remark, the Noether Theorem states that conserved
quantities are related to the existence of cyclic variables into
dynamics \cite{arnold,marmo,morandi}.

It is worth noticing that the Hessian determinant of  Eq. (\ref{lag2}),
${\displaystyle \left|\left|\frac{\partial^2\mathcal{L}}{\partial
q'_i\partial q'_j}\right|\right|}$, is zero. This result clearly
depends on the absence of the generalized velocity $B'$ into the
point\,-\,like Lagrangian. As matter of fact, using a point-like
Lagrangian approach implies that the metric variable $B$ does not
contributes to dynamics, but the  equation of motion for $B$ has
to be considered as a further constraint equation. Then  the
Lagrangian (\ref{lag2})  has three  degrees of freedom and not
four, as one should expect  {\it a priori}.

Now, since the  equation of motion describing the evolution of the
metric potential $B$ does not depend on its derivative, it can be
explicitly solved in term of $B$ as a function of the other
coordinates\,:
\begin{equation}\label{eqb}
B=\frac{2M^2f''A'R'+2Mf'A'M'+4AMf''M'R'+Af'M'^2}{2AM[(2+MR)f'-Mf]}\,.
\end{equation}
By inserting Eq.(\ref{eqb}) into the Lagrangian (\ref{lag2}),  we
obtain a non-vanishing Hessian matrix removing the singular
dynamics. The new Lagrangian reads\footnote{Lowering  the
dimension of configuration space through the substitution
(\ref{eqb}) does not affect the  dynamics since $B$ is a
non-evolving quantity. In fact, inserting Eq. (\ref{eqb})
 into the dynamical equations given by (\ref{lag2}),  they
coincide with those derived by (\ref{lag2}).}
\begin{equation}\label{}\mathcal{L}^*= {\bf L}^{1/2}\end{equation}
with
\begin{eqnarray}\label{lag3}\nonumber
{\bf L}=\underline{q'}^t\hat{{\bf
L}}\underline{q'}=\frac{[(2+MR)f'-fM]}{M}[2M^2f''A'R'
+2MM'(f'A'+2Af''R') +Af'M'^2]\,.
\end{eqnarray}

If one assumes the spherical
symmetry, the role of the {\it affine parameter} is  played by the
coordinate radius $r$.  In this case, the configuration space is
given by $\mathcal{Q}=\{A, M, R\}$ and the tangent space by
$\mathcal{TQ}=\{A, A', M, M', R, R'\}$. On the other hand,
according to the Noether Theorem, the existence of a symmetry for
dynamics described by  Lagrangian (\ref{lag2})  implies a
constant of motion. Let us apply the Lie derivative to the
(\ref{lag2}), we have\footnote{From now on,  $\underline{q}$
indicates the vector $\{A,M,R\}$.}\,:
\begin{equation}\label{}
L_{\mathbf{X}}{\bf L}\,=\,\underline{\alpha}\cdot\nabla_{q}{\bf L
}+\underline{\alpha}'\cdot\nabla_{q'}{\bf L}
=\underline{q}'^t\biggl[\underline{\alpha}\cdot\nabla_{q}\hat{{\bf
L }}+ 2\biggl(\nabla_{q}\alpha\biggr)^t\hat{{\bf L
}}\biggr]\underline{q}'\,,
\end{equation}
that vanishes if the functions ${\underline{\alpha}}$ satisfy the
following system
\begin{equation}\label{sys}
\underline{\alpha}\cdot\nabla_{q}\hat{{\bf L}}
+2(\nabla_{q}{\underline{\alpha}})^t\hat{{\bf L
}}\,=\,0\,\longrightarrow\ \ \ \ \alpha_{i}\frac{\partial
\hat{{\bf L}}_{km}}{\partial
q_{i}}+2\frac{\partial\alpha_{i}}{\partial q_{k}}\hat{{\bf L
}}_{im}=0\,.
\end{equation}
Solving the system (\ref{sys})  means to find out the functions
$\alpha_{i}$ which assign the Noether vector \cite{arnold,marmo}. However the system
(\ref{sys}) implicitly depends on the form of $f(R)$ and then, by
solving it, we get also $f(R)$-models compatible with spherical
symmetry. On the other hand, by choosing the $f(R)$-form, we can
explicitly solve (\ref{sys}). As an example, one finds that the
system (\ref{sys}) is satisfied if we choose
\begin{equation}\label{solsy}f(R)\,=\,f_0 R^s\,,\ \ \ \ \  \underline{\alpha}=(\alpha_1,\alpha_2,\alpha_3)=
\biggl((3-2s)kA,\ -kM,\ kR\biggr)\,,
\end{equation}
with $s$ a real number, $k$ an integration constant and $f_0$ a
dimensional coupling constant\footnote{ The dimensions are given
by $R^{1-s}$ in terms of the Ricci scalar. For the sake of
simplicity, we will put $f_0=1$ in the forthcoming discussion.}.
This means that, for any $f(R)=R^s$, exists, at least, a Noether
symmetry and  a related constant of motion $\Sigma_{0}$\,:
\begin{eqnarray}\label{cm}\nonumber
\Sigma_{0}\,=\,\underline{\alpha}\cdot\nabla_{q'}{\bf L
}=2
skMR^{2s-3}[2s+(s-1)MR][(s-2)RA'-(2s^2-3s+1)AR']\,.\end{eqnarray}
A physical interpretation of $\Sigma_{0}$  is possible if one
gives an interpretation of this quantity in GR, that means for $f(R)=R$ and $s=1$. In other words, the above procedure has to be applied to the
Lagrangian of GR. We obtain the solution
\begin{equation}\label{solsygr}\underline{\alpha}_{GR}=(-kA,\
kM)\,.
\end{equation}
The functions $A$ and $M$ give the Schwarzschild solution, and
then the constant of motion acquires the standard form
\begin{equation}\label{cmgr}\Sigma_{0}= \frac{2GM}{c^2}\,.\end{equation}
In other words, in the case of Einstein gravity, the Noether
symmetry gives, as a conserved quantity,   the Schwarzschild radius
or  the mass of the gravitating system.  This result can be assumed as a consistency check.

In the general case, $f(R)=R^s$, the Lagrangian (\ref{lag2})
becomes
\begin{eqnarray}\label{}\nonumber {\bf L}=\frac{sR^{2s-3}[2s+(s-1)MR]}{M}[2(s-1)M^2A'R'+2MRM'A'+4(s-1)AMM'R'+ARM'^2]\,,\end{eqnarray}
and the expression (\ref{eqb}) for $B$ is
\begin{equation}\label{}B=\frac{s[2(s-1)M^2A'R'+2MRM'A'+4(s-1)AMM'R'+ARM'^2]}{2AMR[2s+(s-1)MR]}\end{equation}
As it can be easily checked, GR is recovered for $s=1$.

Using the constant of motion (\ref{cm}),  we solve in term of $A$
and obtain
\begin{equation}\label{}A=R^{\frac{2s^2-3s+1}{s-2}}\biggl\{k_1+\Sigma_{0}\int\frac{R^{\frac{4s^2-9s+5}{2-s}}dr}{2ks(s-2)M[2s+(s-1)MR]}\biggr\}\end{equation}
for $s\neq2$ and  $k_1$ an integration constant. For $s\,=\,2$,
one finds
\begin{equation}\label{}A=-\frac{\Sigma_{0}}{12kr^2(4+r^2R)RR'}\,.\end{equation}
These relations allow to find out general solutions for the field
equations giving the dependence of the Ricci scalar on the radial
coordinate $r$. For example, a solution is found for
\begin{equation}\label{}
s=5/4\,,\ \ \ \ M=r^2\,,\ \ \ \ R= 5 r^{-2}\,,
\end{equation}
obtaining  the spherically symmetric space-time
\begin{equation}\label{sol_noe_2}ds^2=(\alpha+\beta
r)dt^2-\frac{1}{2}\frac{\beta r}{\alpha+\beta r}
dr^2-r^2d\Omega\,,\end{equation}
where $\alpha$ is a combination of  $\Sigma_0$ and $k$ and $\beta=k_1$.
In principle, the same procedure can be worked out any time Noether symmetries are identified.
Our task is now to show how, from  a spherically symmetric solution, one can  generate an axially symmetric solution adopting the Newman-Janis procedure that works in GR.  In general, the approach is not immediately straightforward since, as soon as $f(R)\neq R$, we are dealing with fourth-order field equations which have, in principle, different existence theorems and boundary conditions. However, the existence of the Noether symmetry guarantees the consistency of the chosen $f(R)$-model with the field equations.

\section{Axial symmetry derived from spherically symmetric solutions}
\label{tre}
We want to show now how it is possible to obtain an axially symmetric solution starting from a spherically symmetric one adopting the method developed by Newman and Janis in GR. Such an algorithm can be applied to a
static spherically symmetric  metric considered as a``seed'' metric.
Let us recast   the  spherically symmetric metric (\ref{me}) in the form

\begin{equation}
ds^2 = e^{2\phi(r)}dt^2 - e^{2\lambda(r)}dr^2 - r^2d\Omega,
\label{eqn:ssm}
\end{equation}
with $g_{tt}(t,r)\,=\,e^{2\phi(r)}$ and
$g_{rr}(t,r)\,=\,e^{2\lambda(r)}$. Such a form is suitable for the considerations below. Following Newman and Janis,
Eq.~(\ref{eqn:ssm}) can be written in the so called
Eddington--Finkelstein coordinates $(u,r,\theta,\phi)$, i.e. the
$g_{rr}$ component is eliminated by a change of coordinates and a
cross term is introduced \cite{gravitation}. Specifically this is achieved by defining
the time coordinate as $dt = du + F(r)dr$ and setting $F(r)= \pm
e^{\lambda(r)-\phi(r)}$. Once such a transformation is performed,  the metric
(\ref{eqn:ssm})  becomes

\begin{equation}\label{nullelemn}
ds^2 = e^{2\phi(r)}du^2 \pm 2e^{\lambda(r)+\phi(r)}dudr -
r^2d\Omega.
\end{equation}
The surface $u\,=\,$ costant is a light cone starting
from the origin $r\,=\,0$. The metric tensor for the line element
(\ref{nullelemn}) in null-coordinates is

\begin{equation}\label{metrictensorcontro}
g^{\mu\nu} = \left(
\begin{array}{cccc}
0 & \pm e^{-\lambda(r)-\phi(r)} & 0 & 0     \\
\pm e^{-\lambda(r)-\phi(r)} & -e^{-2\lambda(r)} & 0 & 0 \\
0 & 0 & -1/r^2 & 0 \\
0 & 0 & 0 & -1/(r^2\sin^2{\theta})
\end{array}    \right).
\end{equation}
The matrix  (\ref{metrictensorcontro}) can be written in terms of a null
tetrad as

\begin{equation}\label{eq:gmet}
g^{\mu\nu} = l^\mu n^\nu + l^\nu n^\mu-m^\mu\bar{m}^\nu -
m^\nu\bar{m}^\mu,
\end{equation}
where $l^\mu$, $n^\mu$, $m^\mu$ and $\bar{m}^\mu$ are the vectors
satisfying the conditions

\begin{equation}l_\mu l^\mu\,=\,m_\mu m^\mu\,=\,n_\mu n^\mu\,=\,0,\,\,\,\,\,
l_\mu n^\mu\,=\,-m_\mu\bar{m}^\mu\,=\,1, \,\,\,\,\,l_\mu
m^\mu\,=n_\mu m^\mu\,=\,0\,.\end{equation} The bar indicates the
complex conjugation. At any point in space, the tetrad can be chosen
in the following manner: $l^\mu$ is the outward null vector
tangent to the cone, $n^\mu$ is the inward null vector pointing
toward the origin, and $m^\mu$ and $\bar{m}^\mu$ are the vectors
tangent to the two-dimensional sphere defined by constant $r$ and
$u$. For the spacetime (\ref{metrictensorcontro}), the tetrad null
vectors can be

\begin{equation}
\left\{\begin{array}{ll}l^\mu\,=\,\delta^\mu_1\\\\n^\mu\,=\,-\frac12
e^{-2\lambda(r)}\delta^\mu_1+
e^{-\lambda(r)-\phi(r)}\delta^\mu_0\\\\
m^\mu\,=\,\frac{1}{\sqrt{2}r}(\delta^\mu_2
+\frac{i}{\sin{\theta}}\delta^\mu_3)\\\\
\bar{m}^\mu\,=\,\frac{1}{\sqrt{2}r}(\delta^\mu_2
-\frac{i}{\sin{\theta}}\delta^\mu_3)\end{array}\right.
\end{equation}
Now we need to extend the set of coordinates
$x^\mu\,=\,(u,r,\theta,\phi)$ replacing the real
radial coordinate by a complex variable. Then the tetrad
null vectors become \footnote{It is worth noticing that a certain
arbitrariness  is present in the
complexification process of the functions $\lambda$ and $\phi$.  Obviously, we have to obtain the metric (\ref{metrictensorcontro})  as soon as $r\,=\,\bar{r}$.}

\begin{equation}\label{tetradvectors}
\left\{\begin{array}{ll}l^\mu\,=\,\delta^\mu_1\\\\n^\mu\,=\,-\frac12
e^{-2\lambda(r,\bar{r})}\delta^\mu_1+
e^{-\lambda(r,\bar{r})-\phi(r,\bar{r})}\delta^\mu_0\\\\
m^\mu\,=\,\frac{1}{\sqrt{2}\bar{r}}(\delta^\mu_2
+\frac{i}{\sin{\theta}}\delta^\mu_3)\\\\
\bar{m}^\mu\,=\,\frac{1}{\sqrt{2}r}(\delta^\mu_2
-\frac{i}{\sin{\theta}}\delta^\mu_3)\end{array}\right.
\end{equation}
A new metric is obtained by making a complex coordinates
transformation

\begin{equation}\label{transfo}x^\mu\rightarrow\tilde{x}^\mu=x^\mu+iy^\mu(x^\sigma)\,,\end{equation}
where $y^\mu(x^\sigma)$ are analityc functions of the real
coordinates $x^\sigma$, and simultaneously let the null
tetrad vectors $Z^\mu_a\,=\,(l^\mu,n^\mu,m^\mu,\bar{m}^\mu)$, with
$a\,=\,1,2,3,4$, undergo the transformation

\begin{equation}Z^\mu_a\rightarrow \tilde{Z}^\mu_a(\tilde{x}^\sigma,\bar{\tilde{x}}^\sigma)\,=\,Z^\rho_a\frac{\partial\tilde{x}
^\mu}{\partial x^\rho}.\end{equation} Obviously, one has to recover the old tetrads and metric as soon as
$\tilde{x}^\sigma\,=\,\bar{\tilde{x}}^\sigma$. In summary, the
effect of the "\emph{tilde transformation}" (\ref{transfo}) is to
generate a new metric whose components are (real) functions of
complex variables, that is

\begin{equation}g_{\mu\nu}\rightarrow \tilde{g}_{\mu\nu}\,:\,\tilde{\mathbf{x}}\times\tilde{\mathbf{x}}\mapsto
\mathbb{R}\end{equation} with

\begin{equation}\tilde{Z}^\mu_a(\tilde{x}^\sigma,\bar{\tilde{x}}^\sigma)|_{\mathbf{x}=\tilde{\mathbf{x}}}
=Z^\mu_a(x^\sigma).\end{equation} For our aims,  we can make
 the choice

\begin{equation}\label{transfo_1}
\tilde{x}^\mu\,=\,x^\mu+ia(\delta^\mu_1-\delta^\mu_0)\cos\theta\rightarrow
\left\{\begin{array}{ll}\tilde{u}\,=\,u+ia\cos\theta\\\\\tilde{r}\,=\,r-ia\cos\theta\\\\
\tilde{\theta}\,=\,\theta\\\\
\tilde{\phi}\,=\,\phi\\\\
\end{array}\right.\end{equation}
where $a$ is constant and the tetrad null vectors
(\ref{tetradvectors}), if we choose
$\tilde{r}\,=\,\bar{\tilde{r}}$, become

\begin{equation}\label{tetradvectors_2}
\left\{\begin{array}{ll}\tilde{l}^\mu\,=\,\delta^\mu_1\\\\\tilde{n}^\mu\,=\,-\frac12
e^{-2\lambda(\tilde{r},\theta)}\delta^\mu_1+
e^{-\lambda(\tilde{r},\theta)-\phi(\tilde{r},\theta)}\delta^\mu_0\\\\
\tilde{m}^\mu\,=\,\frac{1}{\sqrt{2}(\tilde{r}-ia\cos\theta)}\biggl[ia(\delta^\mu_0-\delta^\mu_1)\sin\theta+\delta^\mu_2
+\frac{i}{\sin{\theta}}\delta^\mu_3\biggr]\\\\
\bar{\tilde{m}}^\mu\,=\,\frac{1}{\sqrt{2}(\tilde{r}+ia\cos\theta)}\biggl[-ia(\delta^\mu_0-\delta^\mu_1)\sin\theta+\delta^\mu_2
-\frac{i}{\sin{\theta}}\delta^\mu_3\biggr]\end{array}\right.
\end{equation}

From the transformed null tetrad vectors, a new metric is recovered
using (\ref{eq:gmet}). For the null tetrad vectors given by
(\ref{tetradvectors_2}) and the transformation given by
(\ref{transfo_1}), the new metric, with coordinates
$\tilde{x}^\mu\,=\,(\tilde{u},\tilde{r},\theta,\phi)$, is

\begin{equation}
\tilde{g}^{\mu\nu} = \left(
\begin{array}{cccc} \label{eqn:newmetric}
-\frac{a^2 \sin^2{\theta}}{\Sigma^2} &
e^{-\lambda(\tilde{r},\theta)-\phi(\tilde{r},\theta)} +
\frac{a^2\sin^2{\theta}}{\Sigma^2} & 0 & -\frac{a}{\Sigma^2} \\
. & - e^{-2\lambda(\tilde{r},\theta)} -
\frac{a^2\sin^2{\theta}}{\Sigma^2}
& 0 & \frac{a}{\Sigma^2} \\
. & . & -\frac{1}{\Sigma^2} & 0 \\
. & . & . & -\frac{1}{\Sigma^2\sin^2{\theta}} \\
\end{array}      \right)
\end{equation}
where $\Sigma = \sqrt{\tilde{r}^2 + a^2\cos^2{\theta}}$. In the
covariant form,  the metric (\ref{eqn:newmetric}) is

\begin{equation} \label{eqn:coform}
\tilde{g}_{\mu\nu} =      \left(
\begin{array}{cccc}
e^{2\phi(\tilde{r},\theta)} & e^{\lambda(\tilde{r},\theta) +
\phi(\tilde{r},\theta)} & 0 & a
e^{\phi(\tilde{r},\theta)}[e^{\lambda(\tilde{r},\theta)}-
e^{\phi(\tilde{r},\theta)}] \sin^2{\theta} \\ . & 0 & 0 & -a e^{\phi(\tilde{r},\theta)+\lambda(\tilde{r},\theta)}\sin^2{\theta} \\
. & . & -\Sigma^2 & 0 \\
. & . & . & -[\Sigma^2 +
a^2\sin^2{\theta}e^{\phi(\tilde{r},\theta)}(2e^{\lambda(\tilde{r},\theta)}-
e^{\phi(\tilde{r},\theta)})]\sin^2{\theta} \\
\end{array}      \right)
\end{equation}
Since the metric is symmetric, the dots in the matrix  are used to indicate
$g^{\mu\nu} = g^{\nu\mu}$. The form of this metric gives the
general result of the Newman-Janis algorithm starting from  any spherically
symmetric "seed" metric.

The metric given in Eq.~(\ref{eqn:coform}) can be  simplified by a further gauge  transformation so that the only off-diagonal
component is $g_{\phi t}$. This procedure makes it easier to compare with
the standard Boyer-Lindquist form of the Kerr metric
\cite{gravitation} and to interpret physical properties such as the frame
dragging. The coordinates $\tilde{u}$ and $\phi$ can be
redefined in such a way that the metric in the new coordinate
system has the properties described above. More explicitly, if we
define the coordinates in the following way

\begin{equation}d\tilde{u}\,=
\,dt+g(\tilde{r})d\tilde{r}\,\,\,\,\,\text{and}\,\,\,\,\,d\phi\,=\,d\phi+h(\tilde{r})d\tilde{r}\end{equation}
where

\begin{equation}
\left\{\begin{array}{ll}g(\tilde{r})=-\frac{e^{\lambda(\tilde{r},\theta)}
(\Sigma^2+a^2\sin^2{\theta}e^{\lambda(\tilde{r},\theta)+\phi(\tilde{r},\theta)})}
{e^{\phi(\tilde{r},\theta)}(\Sigma^2+a^2\sin^2{\theta}e^{2\lambda(\tilde{r},\theta)})}\\\\h(\tilde{r})
=-\frac{ae^{2\lambda(\tilde{r},\theta)}}{\Sigma^2+a^2\sin^2{\theta}e^{2\lambda(\tilde{r},\theta)}}\end{array}\right.
\end{equation}
after some algebraic manipulations, one finds that, in
$(t,\tilde{r},\theta,\phi)$ coordinates system, the metric
(\ref{eqn:coform}) becomes
\begin{equation} \label{eq:blform}
g_{\mu\nu} =  \left(
\begin{array}{cccc}
e^{2\phi(\tilde{r},\theta)} & 0 & 0 & a
e^{\phi(\tilde{r},\theta)}[e^{\lambda(\tilde{r},\theta)}-
e^{\phi(\tilde{r},\theta)}] \sin^2{\theta} \\
. & -\Sigma^2/(\Sigma^2 e^{-2\lambda(\tilde{r},\theta)} +
a^2\sin^2{\theta})
 & 0 & 0 \\
. & . & -\Sigma^2 & 0 \\
. & . & . & -[\Sigma^2 +
a^2\sin^2{\theta}e^{\phi(\tilde{r},\theta)}(2e^{\lambda(\tilde{r},\theta)}-
e^{\phi(\tilde{r},\theta)})]\sin^2{\theta} \\
\end{array} \right).
\end{equation}

This metric represents the complete family of
metrics that may be obtained by performing the Newman-Janis
algorithm on any static spherically symmetric "seed" metric, written
in Boyer-Lindquist type coordinates. The validity of these
transformations requires the condition $\Sigma^2+a^2\sin^2\theta
e^{2\lambda(\tilde{r},\theta)}\neq 0$, where  $e^{2\lambda(\tilde{r},\theta)}> 0$. Our task is now to show that such an approach can be used to derive axially symmetric solutions also in $f(R)$-gravity.

\section{Axially symmetric solutions in $f(R)$-gravity: an example}
\label{quattro}

Starting from the above spherically symmetric solution (\ref{sol_noe_2}),
 the metric tensor, written in the Eddington--Finkelstein
 coordinates $(u,r,\theta,\phi)$
of the form (\ref{metrictensorcontro}) is

\begin{equation}\label{metrictensorcontro_noe}
g^{\mu\nu} = \left(
\begin{array}{cccc}
0 & \sqrt{\frac{2}{\beta r}} & 0 & 0     \\
. & -2-\frac{2\alpha}{\beta r}& 0 & 0 \\
. & . & -1/r^2 & 0 \\
. & . & . & -1/(r^2\sin^2{\theta})
\end{array}    \right).
\end{equation}
The complex  tetrad  null vectors (\ref{tetradvectors}) are now

\begin{equation}\label{tetradvectors_noe}
\left\{\begin{array}{ll}l^\mu\,=\,\delta^\mu_1\\\\n^\mu\,=\,-
\biggl[1+\frac{\alpha}{\beta}\biggl(\frac{1}{\bar{r}}+\frac{1}{r}\biggr)\biggr]\delta^\mu_1+
\sqrt{\frac{2}{\beta}}\frac{1}{\sqrt[4]{\bar{r}r}}\delta^\mu_0\\\\
m^\mu\,=\,\frac{1}{\sqrt{2}\bar{r}}(\delta^\mu_2
+\frac{i}{\sin{\theta}}\delta^\mu_3)\end{array}\,.\right.
\end{equation}

By computing the complex coordinates transformation
(\ref{transfo_1}),  the tetrad null vectors become

\begin{equation}\label{tetradvectors_noe_1}
\left\{\begin{array}{ll}\tilde{l}^\mu\,=\,\delta^\mu_1\\\\\tilde{n}^\mu\,=\,-
\biggl[1+\frac{\alpha}{\beta}\frac{\text{Re}\{\tilde{r}\}}{\Sigma^2}\biggr]\delta^\mu_1+
\sqrt{\frac{2}{\beta}}\frac{1}{\sqrt{\Sigma}}\delta^\mu_0\\\\
\tilde{m}^\mu\,=\,\frac{1}{\sqrt{2}(\tilde{r}+ia\cos\theta)}\biggl[ia(\delta^\mu_0-\delta^\mu_1)\sin\theta+\delta^\mu_2
+\frac{i}{\sin{\theta}}\delta^\mu_3\biggr]\end{array}\right.
\end{equation}
Now by performing the same procedure as in previous section,  we derive an axially
symmetric metric of the form (\ref{eq:blform}) but starting from  the spherically symmetric
metric (\ref{sol_noe_2}), that is
\begin{equation}
g_{\mu\nu} =  \left(
\begin{array}{cccc}
\frac{r(\alpha+\beta r)+a^2\beta\cos^2\theta}{\Sigma} & 0 & 0 &
\frac{a(-2\alpha r-2\beta\Sigma^2+\sqrt{2\beta}\Sigma^{3/2})\sin^2\theta}{2\Sigma} \\
. & -\frac{\beta\Sigma^2}{2\alpha r+\beta(a^2+r^2+\Sigma^2)}
 & 0 & 0 \\
. & . & -\Sigma^2 & 0 \\
. & . & . & -\biggl[\Sigma^2 -\frac{a^2(\alpha
r+\beta\Sigma^2-\sqrt{2\beta}\Sigma^{3/2})\sin^2\theta}{\Sigma}\biggr]
\sin^2\theta \\\end{array}\,. \right).
\label{rotating}
\end{equation}

It  is worth noticing that the condition $a=0$ immediately gives  the metric (\ref{sol_noe_2}).  This is nothing else but an example: the method is general and can be extended to any spherically symmetric solution derived in $f(R)$-gravity.
\section{A physical application: geodesics and orbits}
\label{cinque}
Let us  discuss now a physical application of the above result. We will take into account a  freely falling particle moving in the space-time described by the metric  (\ref{rotating}).  For our aims, we
 make explicit use of the Hamiltonian formalism. Given a metric $g_{\mu\nu}$, the motion along
the geodesics  can is described by the Lagrangian
\begin{equation}
{\cal L}(x^{\mu},\dot{x}^\mu)=\dfrac{1}{2}g_{\mu\nu}\dot{x}^{\mu}\dot{x^\nu} \ ,
\end{equation}
where the overdot stands for derivative with respect to an affine parameter
$\lambda$ used to parametrize the curve.  The Hamiltonian description is achieved by considering
the canonical momenta and the Hamiltonian function
\begin{equation}
p_{\mu} = \frac{\de{\cal L}}{\de \dot{x}^{\mu}}=g^{\mu\nu}p_\mu p_\nu \ ,
\;\;\;\;\;\;\;\;\;\;
{\cal H }= p_{\mu}\dot{x}^{\mu}-{\cal L} \ ,
\end{equation}
that results
${\displaystyle
{\cal H}= \frac{1}{2}p_\mu p_\nu g^{\mu\nu} }$.
The advantage of the Hamiltonian formalism with respect to the Lagrangian one is that the resulting
equations of motion do not contain any sign ambiguity coming from turning points  in the orbits (see, for example, \cite{chandra}) .
The Hamiltonian results explicitly independent of time and it is
${\displaystyle
{\cal H}= -\frac{1}{2}m^2 ,}$
where the rest mass $m$ is a constant ($m=0$ for photons).
The geodesic equations are
\begin{equation}
\frac{dx^{\mu}}{d\lambda}=\frac{\partial\cal{ H}}{\partial p_{\mu}}=g^{\mu\nu}p_{\nu}=p^{\mu},
\label{2a}
\end{equation}
\begin{equation}
\frac{dp_{\mu}}{d\lambda}=-\frac{\partial\cal{H}}{\partial x{\mu}}=-\frac{1}{2}\frac{\partial g^{\alpha\beta}}{\partial x^{\mu}}p_{\alpha}p_{\beta}=g^{\gamma\beta}\Gamma^{\alpha}_{\mu\gamma}p_{\alpha}p_{\beta}.
\label{2b}
\end{equation}
In addition, since the Hamiltonian is independent of the affine parameter $\lambda$,
one can directly use the coordinate time as integration parameter.
The problem is so reduced to  solve  six equations of motion.  Using the above definitions, it is easy to achieve the reduced
Hamiltonian (now linear in the momenta)

\begin{equation}
H=-p_{0}=\left[\frac{p_{i}g^{0i}}{g^{00}}+\left[\left(\frac{p_{i}g^{0i}}{g^{00}}\right)^{2}-\frac{m^{2}+p_{i}p_{j}g^{ij}}{g^{00}}\right]^{1/2}\right]\label{eq:e}\end{equation}
with the equations of motion

\begin{equation}
\frac{dx^{i}}{dt}=\frac{\partial H}{\partial p_{i}}\ ,\;\;\;\;\;\;\;\;\;\;\;\;
\frac{dp_{i}}{dt}=-\frac{\partial H}{\partial x^{i}}\ ,\label{eq:4b}\end{equation}
that give the orbits. The method can be applied to the above solution (\ref{rotating}) considering  the following  line element

\begin{eqnarray*}
ds^{2} & = &
\frac{r(\alpha+\beta r)+a^2\beta\cos^2\theta}{\Sigma}dt^2+2\frac{a(-2\alpha r-2\beta\Sigma^2+\sqrt{2\beta}\Sigma^{3/2})\sin^2\theta}{2\Sigma}dtd\phi+\nonumber\\ &&
-\frac{\beta\Sigma^2}{2\alpha r+\beta(a^2+r^2+\Sigma^2)}dr^2 -\Sigma^2 d\theta^2 -\biggl[\Sigma^2 -\frac{a^2(\alpha
r+\beta\Sigma^2-\sqrt{2\beta}\Sigma^{3/2})\sin^2\theta}{\Sigma}\biggr]
\sin^2\theta d\phi^2
 \label{totalmetricnostrpola}\end{eqnarray*}
 by which the elements of the inverse metric can be easily obtained:

\begin{eqnarray}
g^{tt}&=&\frac{4 \Sigma ^2 \left[\Sigma ^2-\frac{a^2 \sin ^2\theta \left(r
   \alpha -\sqrt{2} \sqrt{\beta } \Sigma ^{3/2}+\beta  \Sigma
   ^2\right)}{\Sigma }\right]}{a^2 \sin ^2\theta \left(2 r \alpha
   -\sqrt{2} \sqrt{\beta } \Sigma ^{3/2}+2 \beta  \Sigma ^2\right)^2+4
   \Sigma  \left(a^2 \beta  \cos ^2\theta+r (r \beta +\alpha )\right)
   \left(\Sigma ^2-\frac{a^2 \sin ^2\theta \left(r \alpha -\sqrt{2}
   \sqrt{\beta } \Sigma ^{3/2}+\beta  \Sigma ^2\right)}{\Sigma }\right)}\nonumber\\
   g^{rr}&=&-\frac{\beta  \left(a^2+r^2+\Sigma ^2\right)+2 r \alpha }{\beta  \Sigma
   ^2}\nonumber\\
 g^{\theta\theta}&=&-\frac{1}{\Sigma ^2}\nonumber\\
   g^{t\phi}&=&\frac{2 a \Sigma  \left(-2 r \alpha +\sqrt{2} \sqrt{\beta } \Sigma
   ^{3/2}-2 \beta  \Sigma ^2\right)}{a^2 \sin ^2\theta\left(2 r
   \alpha -\sqrt{2} \sqrt{\beta } \Sigma ^{3/2}+2 \beta  \Sigma
   ^2\right)^2+4 \Sigma  \left[a^2 \beta  \cos ^2\theta +r (r \beta
   +\alpha )\right] \left[\Sigma ^2-\frac{a^2 \sin ^2\theta \left(r
   \alpha -\sqrt{2} \sqrt{\beta } \Sigma ^{3/2}+\beta  \Sigma
   ^2\right)}{\Sigma }\right]}\nonumber\\
     g^{\phi\phi}&=&-\frac{4 \Sigma  \csc ^2\theta \left[a^2 \beta  \cos ^2\theta +r (r
   \beta +\alpha )\right]}{a^2 \sin ^2\theta\left(2 r
   \alpha -\sqrt{2} \sqrt{\beta } \Sigma ^{3/2}+2 \beta  \Sigma
   ^2\right)^2+4 \Sigma  \left[a^2 \beta  \cos ^2\theta +r (r \beta
   +\alpha )\right] \left[\Sigma ^2-\frac{a^2 \sin ^2\theta \left(r
   \alpha -\sqrt{2} \sqrt{\beta } \Sigma ^{3/2}+\beta  \Sigma
   ^2\right)}{\Sigma }\right]}\nonumber\\
  \end{eqnarray}
and the null ones
  \begin{equation}
g^{tr}=g^{t\theta}=  g^{r\theta}=   g^{r\phi}= g^{\theta\phi}=0\,.
   \end{equation}

\begin{figure}
\begin{tabular}{|c|c|}
\hline
\includegraphics[scale = 0.55]{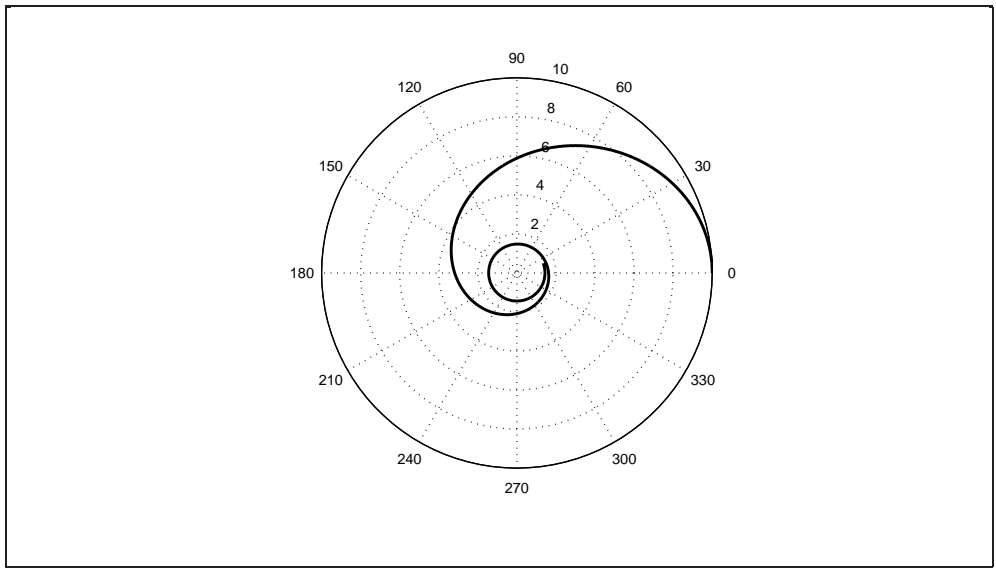}
\tabularnewline
\hline
\end{tabular}
\caption{Relative motion of the test particle with $m=1$.}\label{1}
\end{figure}

Let us consider the equatorial plane, i.e. $\theta=\frac{\pi}{2}$, $\dot{\theta}=0$, and assume $\alpha=1$ and $\beta=2$.
The reduced  Hamiltonian can be written as

\begin{eqnarray}
H(r,\theta, \phi,p_{r},p_{\theta},p_{\phi};t)&=&\frac{2 a p_{\phi} \left(-2 r^3+r^2-1\right)}{a^2 \left(-2 (r-1) r^2-1\right)+r^5}+\left\{\left[ \left(4 a^2 p_{\phi}^{2} \left(-2 r^3+r^2-1\right)^2\right.\right.\right.\nonumber\\ &&
\left.\left.\left.-a^2 \left(-2 (r-1) r^2-1\right)-r^5\right)\left(a^2 \left(r^2 (r (2 r-3) (2 r+1)+6)-2\right)+(2 r+1) r^4\right)\times\right.\right.\nonumber\\&&\left.\left. \left(-\frac{p_{\phi} (2 r+1)}{a^2 \left(r^2 (r (2 r-3) (2 r+1)+6)-2\right)+(2 r+1) r^4}-\frac{p_{r} \left(a^2+r^2+r\right)+p_{\theta}}{r^4}-p_{r}+1\right)
\right]\right\}^\frac{1}{2} .\nonumber\\
  \label{eq:8}
   \end{eqnarray}
It is independent of $\phi$ (i.e. we are considering an azimuthally
symmetric spacetime), and then  the conjugate momentum $p_{\phi}$  is
an  integral of motion.  From Eqs. (\ref{eq:4b}),  one can derive the
coupled equations for $\{r,\theta,\phi,p_{r}$,  $p_{\theta}\}$ and integrate them numerically (the expressions are very cumbersome and will not be reported here). To this goal, we have to
specify the initial value of the position-momentum vector in the phase space.  A  Runge-Kutta method can be used to solve the differential equations. In Fig \ref{1},  the relative trajectories are sketched.

\section{Discussion and Concluding Remarks}
\label{sei}

We have shown that the Newman-Janis method, used  to derive axially symmetric solutions in GR, works also in $f(R)$-gravity.  In principle, it could be consistently applied any time a spherically symmetric solutions is derived. The method does not depend on the field equations but  directly works on the solutions that, a posteriori, has to be checked to fulfill the field equations.

The key point of the method is to find out a suitable  complex transformation which, from a physical viewpoint, corresponds to the fact that we are reducing the number of independent Killing vectors. From a mathematical viewpoint, it is  useful since allows to overcome the problem of a  direct search for  axially symmetric solutions that, in $f(R)$-gravity, could be extremely cumbersome due to the fourth-order  field equations.  However,  other generating techniques  exist and all of them should be explored in order to completely extend solutions of GR to $f(R)$-gravity. They can be  more general and solid than the Newman-Janis approach. A good source for references and basic features of generating techniques is reference \cite{bk:dk}. In particular, the paper by Talbot  \cite{talbot},  considering the Newman-Penrose approach to twisting degenerate metrics, provides some theoretical justification for the scope and limitations of adopting the ``complex trick''.  As reported in Chap. 21 of \cite{bk:dk},  several techniques can be pursued to achieve axially symmetric solutions which can be particularly useful to deal with non-empty space-times ( in particular when perfect fluids are the sources of the field equations) and to deal, in general,  with problems related to  Einstein-Maxwell field equations.
We have to stress that the utility of  generating techniques is not simply to obtain a new metric, but a metric of a new spacetime with specific properties as the transformation properties of the energy-momentum tensor and Killing vectors. In its original application, the Newman--Janis procedure transforms an Einstein-Maxwell solution (Reissner-Nordstrom) into another Einstein-Maxwell solution (Kerr-Newman). As a particular case (setting the charge to zero) it is possible to achieve  the transformation between  two vacuum solutions (Schwarzschild and Kerr).
Also in case of $f(R)$-gravity,  new features emerge by adopting such a technique.  In particular, it is worth studying how certain features  of spherically simmetric metrics, derived in $f(R)$-gravity, result transformed in the new axially symmetric solutions. For example,  considering the $f(R)$  spherically symmetric solution studied here, the Ricci scalar evolves as  $ r^{-2}$ and then the asymptotic flatness is recovered.
Let us consider now the axially symmetric metric achieved by the Newman-Janis method. The parameter $a\neq 0$ indicates that the spherical symmetry ($a=0$) is broken. Such a parameter can be immediately related to the presence of an axis of symmetry and then to the fact that a Killing vector, related to the angle  $\theta$, has been lost.
To conclude, we can say that once the vacuum case is discussed, more general spherical metrics can be transformed in new axially symmetric metrics  adopting more general techniques \cite{bk:dk}. These approaches will be examined and discussed in future works.


\end{document}